\documentclass[twoside,a4paper,11pt]{sca}
\usepackage{graphicx}
\usepackage{hyperref}
\usepackage{movie15}
\begin{document}
\pagenumbering{arabic}
\pagestyle{myheadings}
\thispagestyle{empty}
{\flushright\includegraphics[width=\textwidth,bb=90 650 520 700]{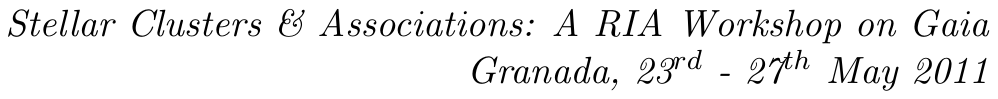}}
\vspace*{0.2cm}
\begin{flushleft}
{\bf {\LARGE
%
On the equivalent effective temperatures of massive young star clusters: The case of NGC 595
%
}\\
\vspace*{1cm}
%
Enrique P\'erez-Montero$^{1}$,
Jos\'e M. V\'\i lchez$^{1}$,
M\'onica Rela\~no$^{2}$,
and
Ana Monreal-Ibero$^{1}$
%
}\\
\vspace*{0.5cm}
%
$^{1}$
Instituto de Astrof\'\i sica de Andaluc\'\i a - CSIC. Apdo. de correos 3004, Granada, Spain\\
$^{2}$
Universidad de Granada, Campus Fuentenueva, Granada, Spain
%
\end{flushleft}
%
\markboth{
On the effective temperature of NGC 595
%
}{ 
%
E. P\'erez-Montero
%
}
\thispagestyle{empty}
\vspace*{0.4cm}
\begin{minipage}[l]{0.09\textwidth}
\ 
\end{minipage}
\begin{minipage}[r]{0.9\textwidth}
\vspace{1cm}
\section*{Abstract}{\small
%
The softness parameter is based on the relative intensity of several optical emission lines emitted by the
gas ionized by young massive star clusters and can be used to derive the equivalent effective temperature ($T_*$) in
those objects whose stellar population cannot be resolved. This method has several uncertainties due to the
disagreement between different synthesis model atmospheres but it is robust to study the relative variations
between objects.

Following the 2D photoionization models of the giant H{\sc ii} region NGC 595 (P\'erez-Montero et al. 2011) 
we show that the determination of $T_*$ with the $\eta$ parameter is also robust in different regions of a same object with large
variations in the geometry of the gas and in the dust-to-gas ratio.
%
\normalsize}
\end{minipage}
%
%
%
\section{Introduction \label{intro}}
The study of the properties of young massive star clusters in our Local Group
can be done by means of the census of the stars that belong to the
corresponding object and the analysis of their properties using a 
colour-magnitude diagram. In the case of those objects whose stellar
population cannot be resolved with the facilities that are nowadays available,
it is necessary to appeal to other tecniques. A possible approach is
to study the massive stars by inspecting their effects on the surrounding
interstellar medium (ISM), including the supply of mechanical energy
from stellar winds and supernovae explosions, the ejection of new metals
produced in the interiors of the stars, or the excitation and
ionization of the atoms in the ISM by the energetic photons emitted by the
stars.
These factors make the optical spectrum of an H{\sc ii}
region to be frequently characterized by the presence of bright prominent 
recombination lines emitted by hydrogen and helium and of collisional lines
emitted by the metals. 
Although the shape of the stellar continuum and some stellar features are
sometimes detectable ({\em e.g.} the Wolf-Rayet features),
the most important source of information about the properties of the
ionizing stellar clusters can only be 
the relative intensities of these lines.

These intensities are mainly
dominated by the so-called functional parameters, including the metallicity ($Z$), the
ionization parameter ($U$, {\em i.e.} the ratio between ionizing photons and the density
of particles), and the equivalent effective temperature ($T_*$). This last is the unique
parameter that only depends on the spectral energy distribution (SED) of the 
ionizing star or cluster.


\section{The {\em softness} parameter}

This parameter, denoted by the greek letter $\eta$, was introduced by
V\'\i lchez \& Pagel (1988) to derive $T_*$ using only
the information from emission-line intensities in the optical spectrum.
It is defined as:

\begin{equation}
\eta = \frac{O^+/O^{2+}}{S^+/S^{2+}} = \frac{[OII] 3727 \AA/[OIII] 4959,5007 \AA}
{[SII] 6717,6731 \AA/[SIII] 9069,9532} + o(Z)
\end{equation}

\noindent depending mainly on the ionic abundances of oxygen and
sulphur ions. This ratio is basically a function of the ratio between the number of ionizing photons 
of O$^+$ (I.P. = 35.1 eV) and S$^+$ (I.P. = 23.3 eV) and has also a certain dependence
on the overall metallicity [$o(Z)$] when it is derived directly from the emission-line intensities.

In the ideal case of a blackbody it only has a linear relation
with the slope of the SED, but for young massive star atmospheres it noticeably
changes for different model atmospheres.
In upper panels of Fig. 1 we show the SEDs of WM-Basic (Pauldrach et al., 2001
with an expanding spherical geometry) at left and
Tlusty (Hubeny \& Lanz, 1995, with a plane-paralel geometry) at right for a same value of $T_*$ (40000 K) and
different values of the metallicity.  The slope and the shape of the 
SEDs can be very different in the energetic range covered by the optical $\eta$ parameter
(Sim\'on-D\'\i az \& Stasi\'nska, 2008).
An alternative to the optical $\eta$ parameter was proposed 
for the mid-infrared spectral range with emission lines of [NeII] (12.8 $\mu$m), [NeIII] (15.6 $\mu$m),
[SIII] (18.7 $\mu$m), and [SIV] (10.5 $\mu$m) (Mart\'\i n-Hern\'andez et al., 2002; Morisset et al. 2004), 
but this involves an even more
energetic range between the ionization potentials of S$^{2+}$ (I.P. = 34.8 eV) and Ne$^⁺$ (I.P. = 40.5 eV),
where discrepancies between the different synthesis model atmospheres are 
larger (see P\'erez-Montero \& V\'\i lchez, 2009, for a more detailed discussion).

In the lower panels of Fig. 1 we show the relation between the optical emission-line ratios
of oxygen ([OII]/[OIII]) and sulphur ([SII]/[SIII]) for a sample of 
objects with emission-line like spectra compiled by P\'erez-Montero \& V\'\i lchez (2009).
The solid lines are quadratical fits to different sets of Cloudy (Ferland et al., 1998) photoionization models with 
different values of $T_*$ (from 35 kK to 50 kK).  At left, these models are calculated using WM-Basic stellar
atmospheres and, at right, Tlusty (see P\'erez-Montero \& V\'\i lchez to see a deeper
descripion of these models).
As can be seen, there
are substantial differences between the two studied synthesis model atmospheres.
These differences prevent an accurate determination of the absolute value of $T_*$
in H{\sc ii} regions but, apparently, allows a good relative characterization 
for different families of objects. In both panels, Circumnuclear Star Forming Regions (CNSFRs)
and H{\sc ii} galaxies have the higher temperatures and, in contrast, smaller regions in our
Galaxy and the Magellanic Clouds have lower temperatures. 

\begin{figure}
\center
\includegraphics[scale=0.3]{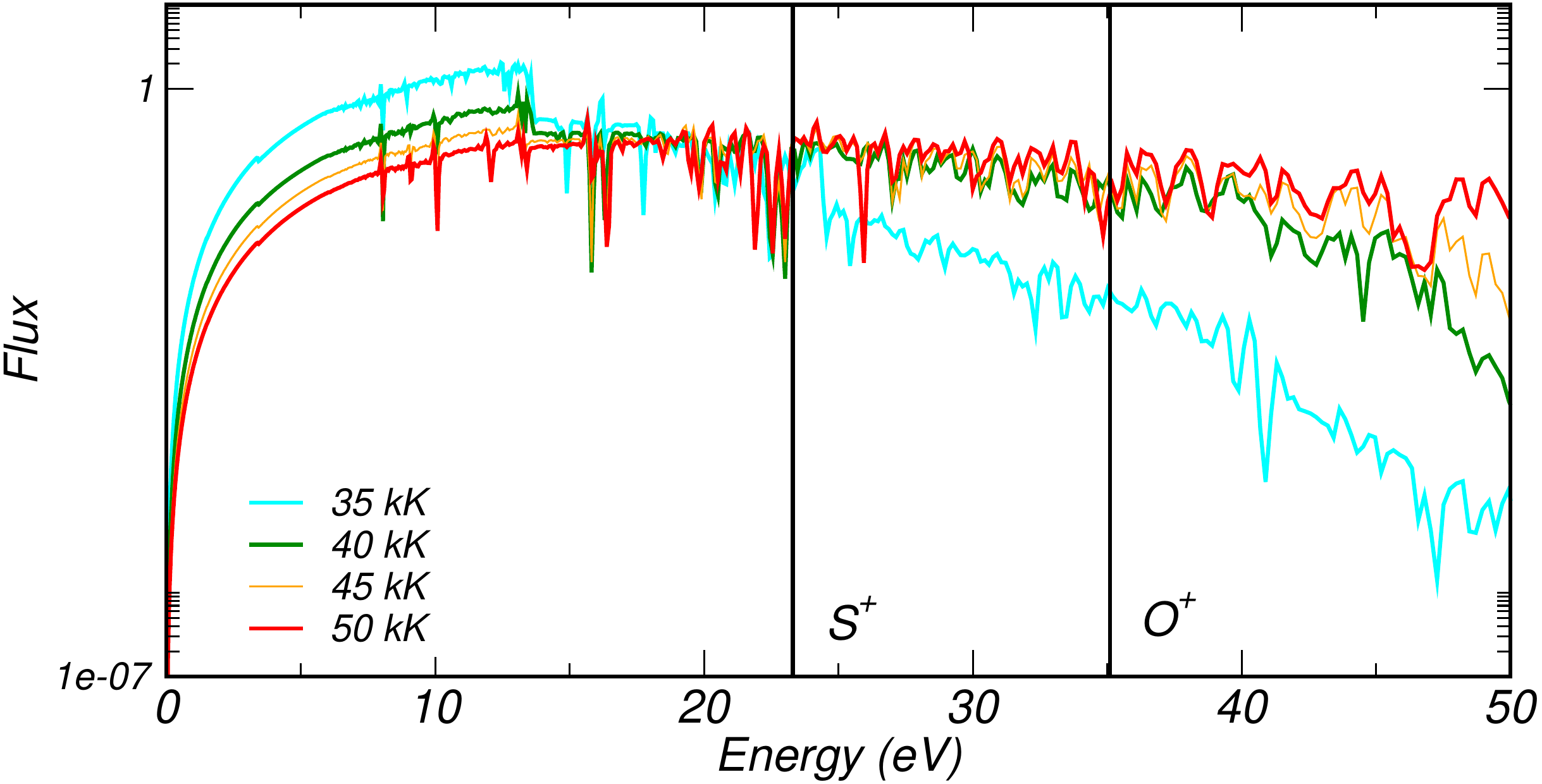} ~
\includegraphics[scale=0.3]{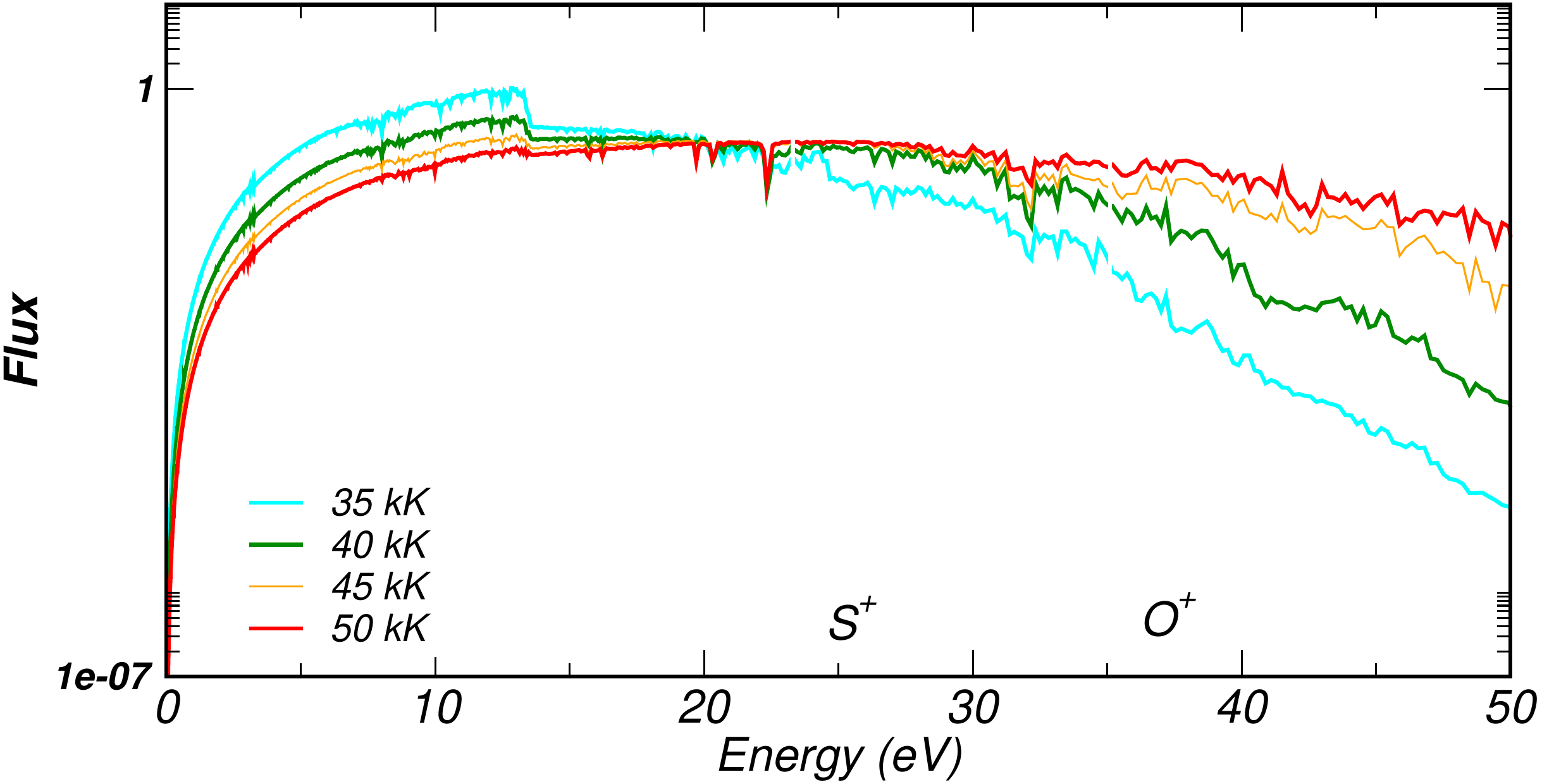} 
\includegraphics[scale=0.43]{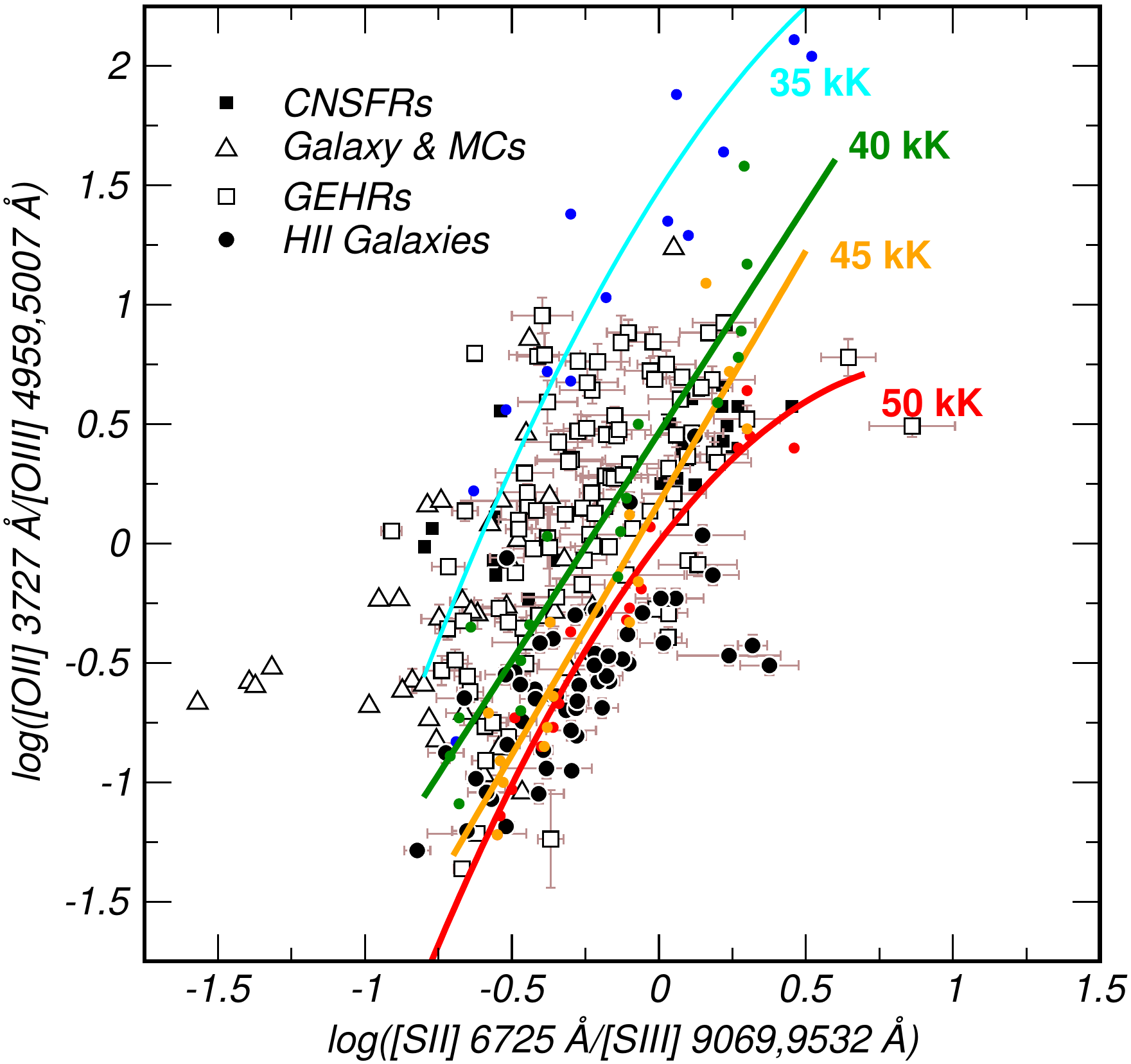}
\includegraphics[scale=0.43]{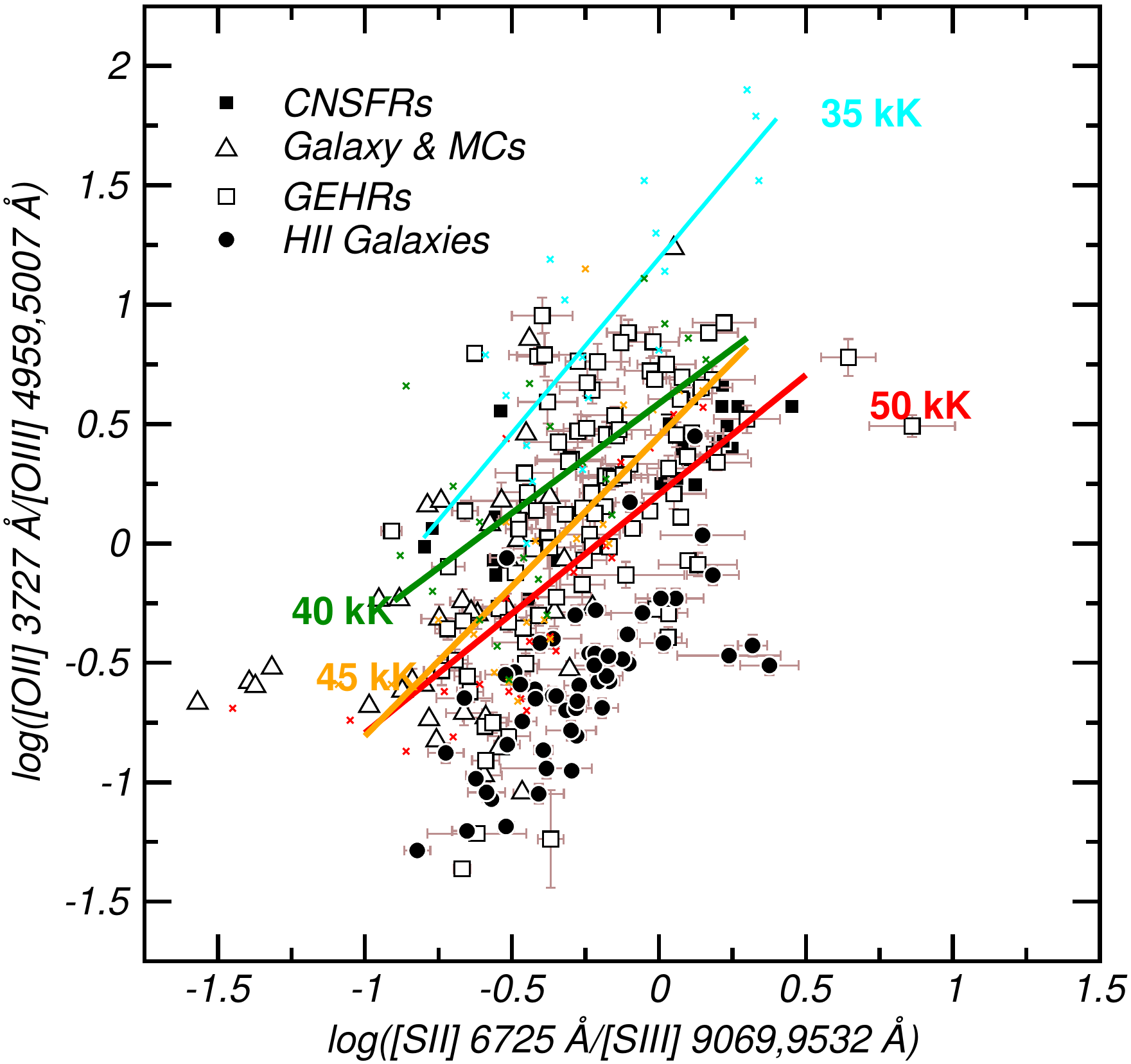} 
\caption{\label{fig1} Upper panels: SEDs of synthesis model atmospheres from
WM-Basic (left) and Tlusty (right) for different values of the metallicity and $T_*$ = 40 kK.
The vertical solid lines represent the ionization potentials of S$^+$ and O$^+$. In the lower panels we show the
relation between the ratios of optical emission line-intensities [OII]/[OIII] and [SII]/[SIII] for different objects.
The solid lines show the quadratical fits to sets of photoionization models with stars of different equivalent effective
temperatures from WM-Basic (left) and Tlusty (right).
} 
\end{figure}

\section{The case of NGC 595}

One ot the possible caveats when using the $\eta$ parameter to find out $T_*$ in
massive star clusters is due to that the information comes from the ionized gas.
This implies that possible variations in the properties of the gas other 
than the functional parameters, such as gas geometry or density, or dust structure could also affect
to the $\eta$ parameter.

We used as a test to check to what extent these variation can affect to the determination
of $T_*$ the Cloudy photoionization models of the 2D structure of the giant H{\sc ii}
region NGC 595 (P\'erez-Montero et al., 2011). The observational
inputs for these models come from integral field spectroscopy taken with the
PMAS instrument at the CAHA 3.5m telescope in the optical range between
3700 {\AA} - 6800 {\AA} (Rela\~no et al., 2010) and photometric information
from Spitzer space observatory at 8 and 24 $\mu$m filters. In these 
models it was assumed a single ionizing source for all the H{\sc ii} region
according to the properties of the CMD diagram derived by Malumuth et al. (1996).
These models reproduce the variations in the optical and mid-IR properties for
different elliptical aperture regions with similar observational properties by 
assuming different matter-bounded geometries combined with different
dust-to-gas ratios, in good agreement to the dust-to-gas ratio
derived in the integrated region (P\'erez-Montero et al.,  2011). In Fig. 2, we show the comparison between the
models and the observations in each elliptical aperture.

\begin{figure}
\center
\includegraphics[scale=0.3]{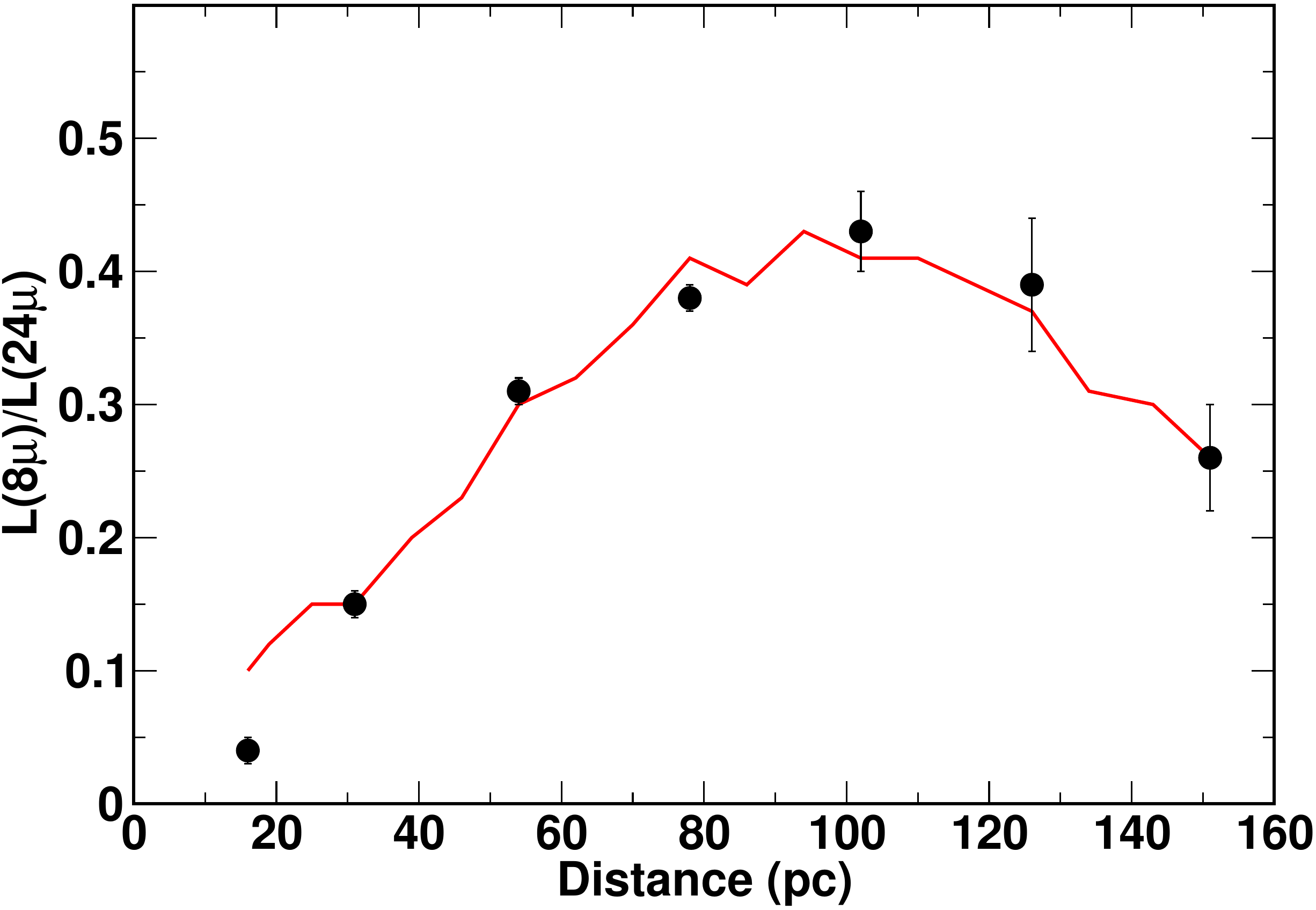} ~
\includegraphics[scale=0.3]{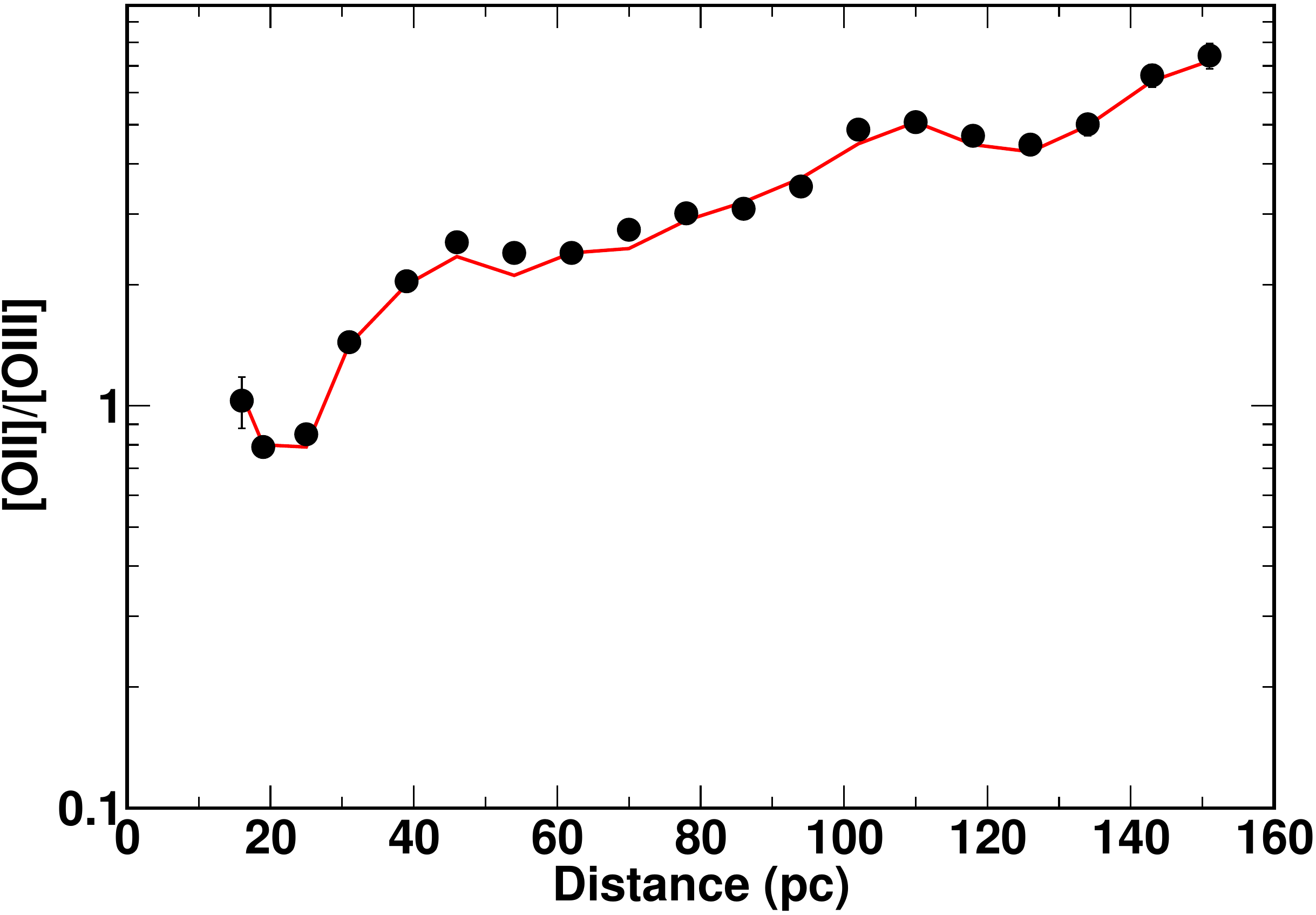} 

\caption{\label{fig2} Variations of the 8 to 24 $\mu$m ratio (at left) and optical [OII]/[OIII] at 
right for different elliptical apertures around the central ionizing cluster in the giant H{\sc ii} region
NGC 595. Black points represent the observed properties and the red solid line the results from
tailored photoionization models for each of these apertures (from P\'erez-Montero et al., 2011)
}
\end{figure}º

In Fig. 3 we show the relation between the two optical emission-line ratios [OII]/[OIII] and
[SII]/[SIII] involved in the $\eta$ parameter with the fits to the photoionization models with Tlusty atmospheres
shown in lower right panel of Fig. 1. The black circles represent the values derived
from the tailored models for each region in NGC 595. As can be seen, with the exception
of some models, the most part of the elliptical apertures lie in regions consistent with a
single value of $T_*$ across the nebula, independently on the inner variations in the
geometry of the gas and in the dust-to-gas ratio. The value of $T_*$ (between
35 kK and 40 kK) is consistent to the value associated with the properties of the cluster
derived by Malumuth using a CMD diagram, which is around 40 kK.
It must be noted, however, that neither the slope of the grid of models for NGC 595 
nor the slopes of the grids of models for different $T_*$ are equal to 1, which
is the value of the slope for the lines with different $\eta$ across the diagram, so this implies
that this parameter must not be used as one-dimensional. In contrast, its
analysis must be undertaken using a 2D approach as in this case. For instance,
in the models of NGC 595, there is a variation in $\eta$ in a factor 2 which is
not contradictory with an homogeneous value of $T_*$ in the models.

\begin{figure}
\center
\includegraphics[scale=0.5]{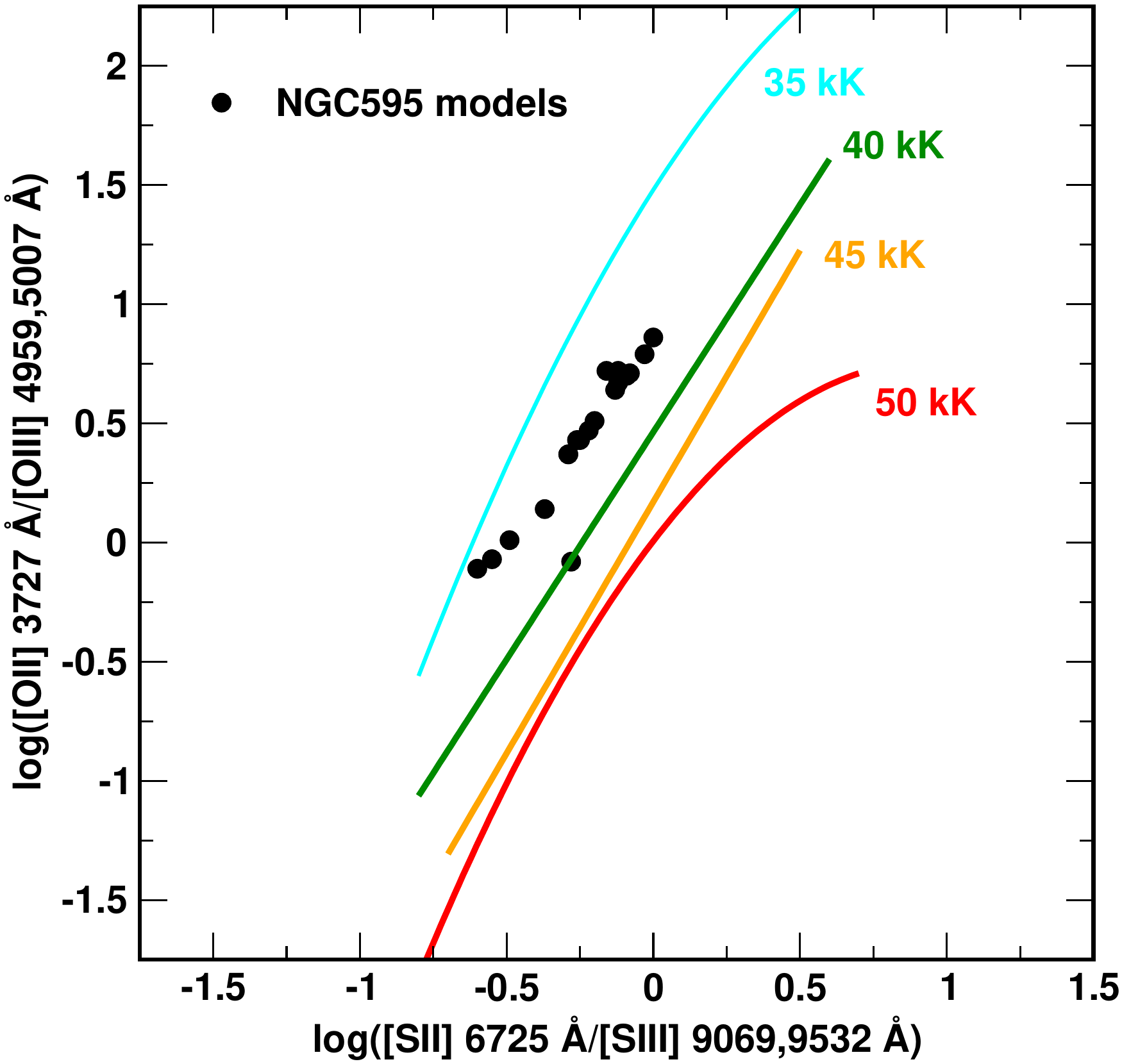}
\caption{\label{fig1} Relation between the optical emission-line ratios [OII]/[OIII] and [SII]/[SIII]. Solid lines
represent the same fits to grids of photoionization models of different $T_*$ for Tlusty atmospheres
represented in Fig. 1. Black circles represent the models for NGC 595 derived by P\'erez-Montero et al. (2011)
}
\end{figure}

\section{Conclusions}
The determination of properties of a massive young stellar cluster can be
obtained by using information from the emission line spectrum from the ionized
gas in those cases where the stellar population of the cluster cannot be resolved.
The $\eta$ parameter as analyzed in 2 dimensions is a robust method to
derive the equivalent effective temperature of a cluster, even in those cases
where there are variations in the geometry of the gas or in the dust-to-gas ratio.
The most important caveat appears as a consequence of the disagreements
in the involved energetic ranges between different synthesis model
atmospheres, but it is possible to use this to study the relative
variations of $T_*$ in homogeneous samples of objects. This is the case of
the variation of $T_*$ across spiral disks found by P\'erez-Montero \& V\'\i lchez (2009),
which points to a hardening of the ionizing radiation above all in spiral galaxies
of low dynamical mass and luminosity with late morphological type.

%
\small  
%
\section*{Acknowledgments}   
%
This work has been supported by the projects AYA2007-
67965-C03-02 of the Spanish National Plan for Astronomy
and Astrophysics and CSD2006 00070 ”1st Science with
GTC” of the Spanish Ministry of Science and Innovation
(MICINN).
%

%

\begin{thebibliography}{}
\small
%
\bibitem{}{Ferland, G.J., Korista, K.T., Verner, D.A., Ferguson, J.W., Kingdom, J.B., \& Verner, E.M.,  1998, PASP, 110, 761}
\bibitem{}{Hubeny, I., \& Lanz, T., 1995, ApJ, 439, 875}
\bibitem{}{Mart\'\i n-Hern\'andez, N. L:, Vermeij, R., Tielens, A.G.G.M., van der Hulst, J.M., \& Peeters, E., 2002, A\&A, 389, 286}
\bibitem{}{Malumuth, E.M., Waller, W.H., \& Parker, J.W., 1996, AJ, 111, 1128}
\bibitem{}{Morisset, C., Schaerer, D., Bouret, J.-C., \& Martins, F., 2004, A\&A, 415, 577}
\bibitem{}{Pauldrach, A.W.A., Hoffmann, T.L., \& Lennon, M., 2001, A\&A, 425, 849}
\bibitem{}{P\'erez-Montero, E., Rela\~no, M., V\'\i lchez, J.M., Monreal-Ibero, A., 2011, MNRAS, 412, 675}
\bibitem{}{P\'erez-Montero, E. \& V\'\i lchez, J.M., 2009, MNRAS, 400, 1721}
\bibitem{}{Rela\~no, M., Monreal-Ibero, V\'\i lchez, J.M., \& Kennicutt, R.C., 2010, MNRAS, 402, 1635}
\bibitem{}{Sim\'on-D\'\i az, S., \& Stasi\'nska, G., 2008, MNRAS, 389, 1009}
\bibitem{}{V\'\i lchez, J.M., Pagel, B.E.J., 1988, MNRAS, 231, 257}
%
%
\end{thebibliography}
\end{document}